\documentclass[pre, twocolumn, showpacs,preprintnumbers,amssymb,superscriptaddress]{revtex4-1}

\usepackage{epsfig, color}
\usepackage{amsthm, amsmath, amsfonts, ae, psfrag}
\usepackage{braket}
\usepackage{bbm}
\usepackage[english]{babel}
\usepackage{epstopdf} 
\usepackage{notes2bib}
\usepackage{verbatim} 

\newcommand{\red}[1]{{\color{black}#1}}

\usepackage[colorlinks,
pdfpagelabels,
pdfstartview = FitH,
bookmarksopen = true,
bookmarksnumbered = true,
linkcolor = blue,
plainpages = false,
hypertexnames = false,
citecolor = blue] {hyperref}

\begin{document}
	
\footnotetext{We found similar good agreement with the asymptotic result in MC simulations for smaller values of the width of the square-well, i.e. $c(r)$ retains the shape and variation with $\eta$ as displayed in Fig.~\ref{Fig:cr_MC}.}

\title{On the decay of the pair correlation function and the line of vanishing excess isothermal compressibility in simple fluids}

\author{Daniel~Stopper}

\email{daniel.stopper@uni-tuebingen.de}

\affiliation{Institute for Theoretical Physics, University of T\"ubingen, Auf der Morgenstelle 14, 72076 T\"ubingen, Germany}
\affiliation{H. H. Wills Physics Laboratory, University of Bristol, Bristol, BS8 1 TL, UK}

\author{Hendrik Hansen-Goos}

\author{Roland Roth}

\email{roland.roth@uni-tuebingen.de}

\affiliation{Institute for Theoretical Physics, University of T\"ubingen, Auf der Morgenstelle 14, 72076 T\"ubingen, Germany}

\author{Robert Evans}

\affiliation{H. H. Wills Physics Laboratory, University of Bristol, Bristol, BS8 1 TL, UK}

\date{\today}

\begin{abstract}
   We re-visit the competition between attractive and repulsive interparticle forces in simple fluids and how this governs and connects the macroscopic phase behavior and structural properties as manifest in pair correlation functions. We focus on the asymptotic decay of the total correlation function $h(r)$ which is, in turn, controlled by the form of the pair direct correlation function $c(r)$. The decay of $r h(r)$ to zero can be either exponential (monotonic) if attraction dominates repulsion and exponentially damped oscillatory otherwise. The Fisher-Widom (FW) line separates the phase diagram into two regions characterized by the two different types of asymptotic decay. We show that there is a new and physically intuitive thermodynamic criterion which approximates well the actual FW line. This new criterion defines a line where the isothermal compressibility takes its ideal gas value $\chi_T=\chi_T^\text{id}$. We test our hypothesis by considering four commonly used models for simple fluids. In all cases the new criterion yields a line in the phase diagram that is close to the actual FW line for the thermodynamic state points that are most relevant. We also investigate (Widom)  lines of maximal correlation length, emphasizing the importance of distinguishing between the true and Ornstein-Zernike correlation lengths.
   
\end{abstract}

\maketitle

\section{Introduction} \label{Sec:Introduction}

The statistical physics of liquids is frequently concerned with the role of repulsive and attractive interparticle potentials, and their competition, in determining the thermodynamic and structural properties. At the most basic level, the virial expansion of the pressure $p = k_B T(\rho_b + B_2(T) \rho_b^2 + \cdots)$ provides a measure of the competition at low number densities $\rho_b$. $T$ is the temperature and $k_B$ is Boltzmann's constant. If repulsion dominates the second virial coefficient is positive, $B_2(T) > 0$, so that the pressure $p$ is larger than the ideal-gas value, $p > p^\text{id} = k_B T \rho_b$, whereas if attraction is dominant then $B_2(T) < 0$ and $p < p^\text{id}$. The Boyle temperature $T_B$, defined by $B_2(T_B)=0$, is that for which repulsive and attractive interactions cancel in a dilute gas. For a Lennard-Jones fluid $k_B T_B / \varepsilon \approx 3.418$ where $\varepsilon$ is the Lennard-Jones well-depth \cite{Vargas2001}. In colloid science and in the physics of proteins the sign and magnitude of the second virial coefficient $B_2$ plays an important role in quantifying the effective interactions between these mesoscopic particles suspended in a solvent \cite{ArcherEvans2003,deHekand1982, Wolf2014}. The value of $B_2$ is also believed to play an important role in determining the onset of phase coexistence in \textit{dense} fluids. The empirical criterion \cite{Lekkerkerker2000, NoroFrenkel2000} for the critical value, i.e. $B_2^\text{crit}/B_2^\text{\tiny HS} \lesssim -1.5$ is often used to estimate the gas-liquid critical temperature. Here $B_2^\text{\tiny HS} = 2\pi \sigma^3/3$ is the second virial coefficient for hard spheres (HS) of diameter $\sigma$. For very short-ranged attractive potentials the adhesive hard-sphere criterion $B_2^\text{crit}/B_2^\text{\tiny HS} \lesssim -1.2$ is preferred \cite{Ashton2011}. These criteria are based on the idea: provided there is sufficient net attraction, as measured by a sufficiently negative $B_2(T)$, phase coexistence can occur.

The competition between repulsive and attractive interatomic forces also governs the form of the pair and higher order correlation functions, i.e. the structure of the fluid. Seminal work \cite{Widom1967, Weeks1971} explained the importance of repulsive forces and their softness in determining the short-ranged behavior of the \textcolor{black}{total correlation function $h(r)\equiv g(r) - 1$, where $g(r)$ is the radial distribution funcion}. Here we focus primarily on the long-ranged behavior of $h(r)$. For a dilute gas at low $T$, or in the vicinity of the gas-liquid critical point, $rh(r)$ should decay to zero exponentially, as $r\rightarrow \infty$; the decay length defines the true correlation length $\xi$. On the other hand, in the liquid state or in a supercritical high density fluid state we expect \textcolor{black}{$rh(r)$} to decay in an exponentially damped oscillatory fashion, similar to the decay found for \textcolor{black}{one-component} HS fluids at all state points. The former mode of asymptotic decay requires sufficient interparticle attraction, whereas the latter is a signature that repulsion is dominating. The crossover between pure exponential and exponentially damped oscillatory decay of \textcolor{black}{$rh(r)$} defines a line in the phase diagram, first identified by Fisher and Widom (FW) \cite{FisherWidom1968} in their analysis of one-dimensional models. They conjectured that similar crossover would occur in three dimensional fluids. Determining the FW line requires knowledge of the poles of the Fourier transform $\widehat{h}(k)$ of the total pair correlation function $h(r)$. In turn, this requires \textcolor{black}{calculating} the pair direct correlation function $c(r)$ \textcolor{black}{at many thermodynamic} state points \cite{HansenMcDonald2013, EvansHenderson2009, Evans1993, Evans1994}. One learns that the form of $c(r)$ is crucial in determining whether the ultimate decay of $r h(r)$ is damped oscillatory or monotonic.

In this paper, we re-visit how the competition between repulsive and attractive interparticle potentials influences the structure of fluids. In particular, we enquire whether there is a simple physical criterion that indicates where in the phase diagram the FW structural crossover should occur. By considering the repulsive and attractive contributions to $c(r)$ we propose a simple approximate criterion: FW crossover should occur close to the line where the isothermical compressibility $\chi_T$ takes its ideal gas value $\chi_T^\text{id}$.

\textcolor{black}{We also investigate the so-called Widom (W) line, which we define as the line of a local maximum of the true correlation length $\xi$. In recent literature, the term \lq Widom line\rq\, is often associated with lines of extrema of thermodynamic response functions, which appears to have its origin in papers from H. E. Stanley and co-workers, see e.g. Ref. \onlinecite{Stanley2005}, dealing with a liquid-liquid transition. In Ref. \onlinecite{Brazhkin2014} several lines of maximal response functions are plotted for the square-well fluid. The title of the paper: \lq True Widom line for a square-well system\rq\, is unfortunate as the authors consider the Ornstein-Zernike (OZ) correlation length $\xi_{OZ}$, which appears in the celebrated expansion of the static structure factor $S(k) = S(0)/(1+\xi_\text{OZ}^2 k^2)$ at low wavenumbers $k\rightarrow0$, \textit{not} the true \cite{Footenote} correlation length $\xi$ which is determined by the asymptotic decay of $r h(r)$}.

 Our paper is arranged as follows: in Sec. \ref{Sec:ParticleForcesAndDecay} we provide background to the FW line and its determination and show how this line describes crossover of the decay of pair correlations. Sec. \ref{Sec:ParticleForcesAndCompr} describes our new conjecture for the importance of $\chi_T = \chi_T^\text{id}$ criterion. In Sec. \ref{Sec:Results} we present results of calculations of the FW and $\chi_T=\chi_T^\text{id}$ lines for four types of model fluid: the square well (SW), the Asakura-Oosawa (AO), the sticky hard sphere, and the hard-core Yukawa models. In all cases we find the two lines lie close in the most physically relevant regions of the phase diagram. We also present results for the Widom line, where the true correlation length $\xi$ is a local maximum. This line emanates from the critical point to higher $T$, through a region of monotonic decay of pair correlations, terminating at the FW line. We conclude in Sec. \ref{Sec:Conclusion} with a discussion of our results.

\section{The FW line and decay of pair correlations} \label{Sec:ParticleForcesAndDecay}

In addition to determining macroscopic phase behavior, i.e. the existence of gaseous, liquid and solid phases, the competition between interparticle attraction and repulsion is also reflected in the microscopic structure of fluids. As mentioned in Sec. \ref{Sec:Introduction},  Fisher and Widom \cite{FisherWidom1968} conjectured that in three-dimensional systems in the fluid phase, the total correlation function $h(r)\equiv g(r)-1$ should decay to zero in damped oscillatory fashion as $r \rightarrow \infty$ at state points for which repulsion dominates over attraction; typically at sufficiently high volume fractions $\eta$ and/or temperatures $T$. In contrast, at state points where attraction dominates, e.g. in proximity to the critical point or in the gaseous phase, $h(r)$ should decay monotonically to zero, from above, as $r\rightarrow\infty$.

The asymptotic behavior of $h(r)$ can be extracted from its Fourier representation along with the Ornstein-Zernike relation \cite{HansenMcDonald2013}. In $d = 3$
\begin{align} \label{Eq:h(r)_Fourierspace}
	r h(r) &= \frac{1}{2\pi^2} \int_{0}^{\infty} \text{d}k\,k\sin(kr)\, \widehat{h}(k) \notag \\
	&= \frac{1}{4\pi^2 i}\int_{-\infty}^{\infty}\text{d}k\,k\,e^{ikr} \frac{\widehat{c}(k)}{1 - \rho_b\widehat{c}(k)}\,,
\end{align}
where $\rho_b = N/V$ is the number density of the fluid and $\widehat{c}(k)$ is the Fourier transform of the bulk pair direct correlation function $c(r)$. Note that the second equation holds only if $\widehat{c}(k)$ is an even function; this is the case for exponentially or faster decaying pair potentials or pair potentials of finite range. If the pair potential decays as a power law, as is the case for dispersion interactions, there are complications \cite{EvansHenderson2009}.  In this paper, we restrict consideration to short-ranged interactions. 

The right-hand side of Eq. \eqref{Eq:h(r)_Fourierspace} can be evaluated by performing a contour integration in the upper half-plane and applying the residue theorem. Provided that all poles of $\widehat{h}(k)$ in the complex plane are simple, it follows that
\begin{equation}
	rh(r) = \sum_{n} e^{i k_n r} A_n\,,
\end{equation}
where $k_n$ is the $n$-th pole satisfying $1 - \rho_b \widehat{c}(k_n) = 0$, and $2 \pi A_n$ is the residue of $q\widehat{c}(k)/(1 - \rho_b\widehat{c}(k))$ at $k = k_n$. Clearly, the pole with the smallest imaginary part determines the asymptotic decay. We term this the leading pole. If this pole is complex it will occur as a conjugate pair: $k_n = \pm \alpha_1 + i\alpha_0^\text{osc}$ and the ultimate decay takes the form
\begin{equation} \label{Eq:AsymptoticDecayh(r)Osc}
	rh(r) \sim \exp(-\alpha_0^\text{osc}\,r)\cos(\alpha_1\,r - \theta)\,,~~r\rightarrow \infty\,,
\end{equation}
where $\theta$ is a phase \cite{Evans1993, Evans1994}. On the other hand, the leading pole may be purely imaginary: $k = i\alpha_0^\text{mon}$ and $\alpha_1 = 0$. Then $r h(r)$ vanishes purely exponentially for $r \rightarrow \infty$, i.e.
\begin{equation} \label{Eq:AsymptoticDecayh(r)Exp}
	rh(r)\sim \exp(-\alpha_0^\text{mon} r)\,,~~r\rightarrow\infty\,.
\end{equation}
In some (approximate) theories one finds leading poles with $\alpha_0^\text{osc} = 0$ and $\alpha_1 > 0$, corresponding to pure oscillatory decay of $r\,h(r)$. These point to an instability of the uniform fluid with respect to density modulations \cite{Henderson1999, Kirkwood1941}. The FW line is the boundary in the phase diagram where pure exponential $(\alpha_1 \equiv 0)$ and damped oscillatory solutions have the same imaginary part, i.e. $\alpha_0^\text{mon} = \alpha_0^\text{osc}$.

For several models and theories $\widehat{c}(k)$ is known analytically so poles of $\widehat{h}(k)$ can be calculated directly. However, in many cases $c(r)$ is only available numerically. Poles can be found by equating real and imaginary parts in the solution to $1 - \rho_b\widehat{c}(k) = 0$. One obtains the following coupled equations \cite{Evans1993, Evans1994} 
\begin{align}
	1 &= 4 \pi\rho_b\int_{0}^{\infty}\text{d}r\,r^2c(r)\frac{\sinh(\alpha_0 r)}{\alpha_0 r} \cos(\alpha_1 r)\,, \label{Eq:DefPolesRealSpace_1}\\
	1 &= 4\pi\rho_b\int_{0}^{\infty}\text{d}r\,r^2c(r)\cosh(\alpha_0 r) \frac{\sin(\alpha_1 r)}{\alpha_1 r}\,. \label{Eq:DefPolesRealSpace_2}
\end{align}
\begin{figure}[t] 
	\centering
	\includegraphics[width = 8cm]{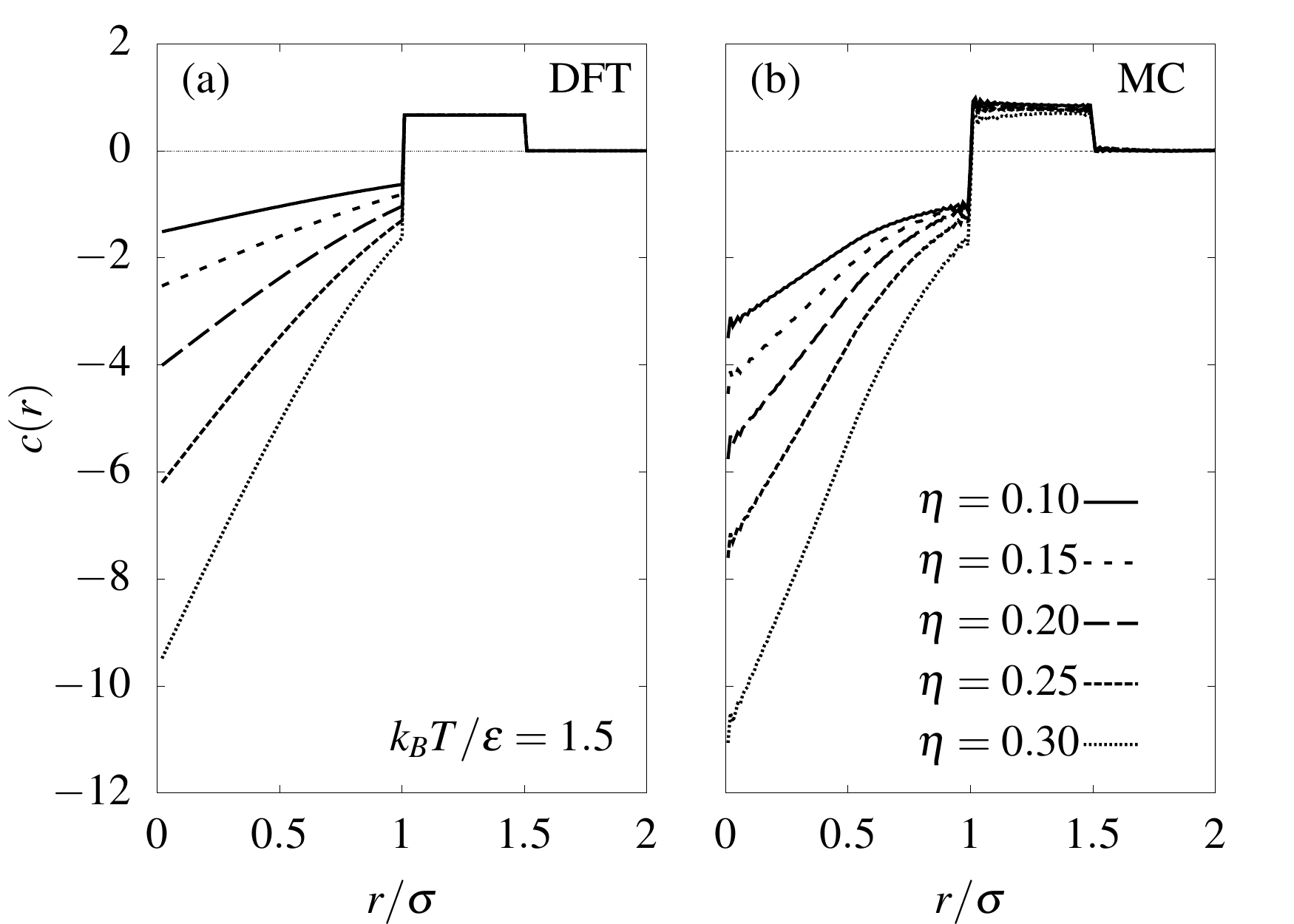}
	\caption{The bulk pair direct correlation function $c(r)$ of the square-well fluid with diameter $\sigma$ and width $0.5 \sigma$ ($\lambda = 1.5$) obtained from (a) a mean-field density functional theory and (b) Monte-Carlo simulations for fixed reduced temperature $k_B T/\varepsilon = 1.5$ and packing fractions $\eta = 0.1-0.3$.} \label{Fig:cr_MC}
\end{figure} 
These equations can be solved numerically to find $\alpha_0$ and $\alpha_1$ for the leading pole at a given state point, provided that the inputted $c(r)$ decays sufficiently quickly to zero so that the integrals converge -- this is typically the case for interparticle potentials decaying faster than a power law.  The leading pure imaginary pole can be found from Eq. \eqref{Eq:DefPolesRealSpace_1} alone with $\alpha_1 = 0$. 
From Eqs. \eqref{Eq:AsymptoticDecayh(r)Osc} and \eqref{Eq:AsymptoticDecayh(r)Exp} we see that $(\alpha_0^\text{mon})^{-1}$ is precisely the (true) correlation length $\xi$ of the fluid. The liquid-gas spinodal corresponds to solutions of Eqs. \eqref{Eq:DefPolesRealSpace_1} \textcolor{black}{and \eqref{Eq:DefPolesRealSpace_2}} with $\alpha_1 = \alpha_0 = 0$ and the FW line is bounded by the liquid spinodal \cite{Evans1993}.

\begin{figure}[t] 
	\centering
	\includegraphics[width = 8cm]{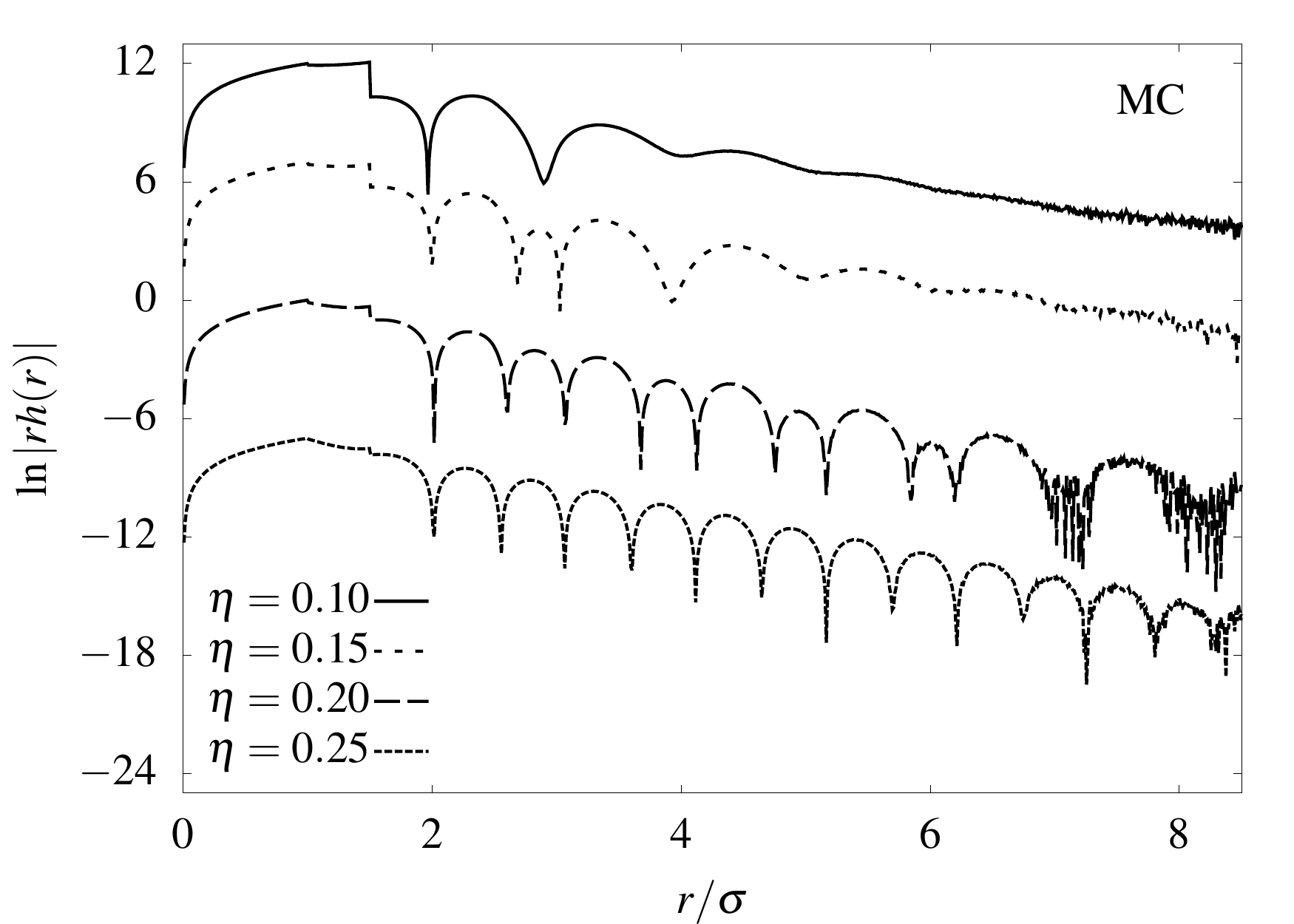}
	\caption{Asymptotic decay of the total correlation function $h(r)$ for the square-well fluid with $\lambda = 1.5$, and fixed $k_B T/\varepsilon = 1.5$ as obtained from MC simulations for the same state points as in Fig. \ref{Fig:cr_MC} apart from $\eta = 0.30$. Fisher-Widom (FW) crossover from monotonic to damped oscillatory decay occurs at roughly $\eta \approx 0.2$. The curves have been shifted vertically for clarity.} \label{Fig:gr_MC}
\end{figure}

It is evident from Eqs. \eqref{Eq:DefPolesRealSpace_1} and \eqref{Eq:DefPolesRealSpace_2} that the location of imaginary poles relative to the complex ones is controlled by the form of the bulk direct correlation function $c(r)$. Typically, in simple fluids, $c(r)$ exhibits a negative repulsive core region for $r < \sigma$, the atomic diameter, arising from repulsive packing effects, and a positive contribution for $r > \sigma$, where the pair potential $\phi(r)$ is attractive \cite{HansenMcDonald2013}. It follows that the asymptotic decay of correlations in a fluid is determined by the competition between repulsive and attractive interactions.
For illustration, in Fig. \ref{Fig:cr_MC} we plot $c(r)$ for a square-well fluid as obtained from (a) a standard mean-field density functional theory \cite{Evans1979, HansenMcDonald2013, ArcherEvans2017} description (more details will be given in Sec. \ref{Sec:Results}) and (b) Monte-Carlo (MC) simulations. In order to extract $c(r)$ from simulation we followed the method described in Ref. \onlinecite{Dijkstra2000}. The reduced temperature $T^* = k_B T/\varepsilon = 1.5$ is fixed well above the critical temperature $T_c^*$, and the packing fraction is $\eta$ is varied. There is good qualitative agreement between theory and simulations: As observed in Ref. \onlinecite{Dijkstra2000} for a truncated and shifted Lennard-Jones fluid, the attractive tail of $c(r)$ is not very sensitive to the packing fraction but the  core contribution \red{becomes more negative} with increasing particle density. Recall \cite{HansenMcDonald2013} that, away from the critical point, $c(r) = -\beta\phi(r)$, $r\rightarrow\infty$. The simulation results  in Fig.~\ref{Fig:cr_MC} (b) show that this asymptotic result remains rather accurate \cite{Note1} down to the core diameter $\sigma$. In Fig.~\ref{Fig:gr_MC} we show the asymptotic decay of $h(r)$ from MC for the same state points as in Fig.~\ref{Fig:cr_MC}. For $\eta = 0.10$ and $0.15$ $rh(r)$ decays monotonically at large $r/\sigma$, consistent with a leading pure imaginary pole. For $\eta = 0.25$ the decay is exponentially damped oscillatory, consistent with a leading conjugate pair of complex poles. The results for $\eta = 0.20$ also point to oscillatory asymptotic decay but this state point lies close to the FW crossover from pure exponential to damped oscillatory decay -- see Sec. \ref{Sec:Results}. Which pole is leading is governed by competition between interparticle attraction and repulsion. However, it is not easy to glean from Eqs. \eqref{Eq:DefPolesRealSpace_1} and \eqref{Eq:DefPolesRealSpace_2} what \textit{physical} criterion determines the location of the FW line. In the next section we seek such a criterion for the crossover. 

\section{Interparticle forces and isothermal compressibility} \label{Sec:ParticleForcesAndCompr}

An insightful paper by Widom \cite{Widom1967} noted that the isothermal compressibility $\chi_T$ can provide a qualitative measure for the overall balance of repulsive and attractive particle interactions in a fluid. We augment his arguments. Recall that $\chi_T$ is related directly \cite{HansenMcDonald2013} to $h(r)$ via the relation
\begin{align} \label{Eq:IsothermalCompressibility}
	\rho_b k_B T \chi_T &=  \left(\frac{\partial \beta p}{\partial \rho_b}\right)_T^{-1} 
	= 1 + 4 \pi\rho_b \int_{0}^{\infty}\text{d}r\,r^2 h(r)\,,
\end{align}
where $\beta = 1/(k_B T)$ denotes the inverse temperature.
Suppose that we are close to the critical point so that pressure gradients $(\partial p/\partial \rho_b)_T$ become very small. Then $\chi_T$ is very large, i.e. $\chi_T/\chi_T^\text{id} \gg 1$, where $\chi_T^\text{id} = (\rho_b k_B T)^{-1}$ is the compressibility of the ideal gas. Such behavior can occur only if $h(r)$ decays monotonically, from above, at large $r$ with a long decay length. Asymptotic decay described by Eq. \eqref{Eq:AsymptoticDecayh(r)Exp} with a positive amplitude, meets this requirement when the correlation length $\xi \equiv (\alpha_0^\text{mon})^{-1}$ is large. The divergence of $\chi_T$ at the critical point, driven by a diverging correlation length $\xi\rightarrow\infty$, requires sufficiently strong attractive interactions so that Eq. \eqref{Eq:AsymptoticDecayh(r)Exp} is valid, i.e. the critical point must lie on the monotonic side of the FW line.

At higher densities or temperatures, where repulsive interactions dominate, $\chi_T$ may fall well below $\chi_T^\text{id}$. Such behavior can occur if $h(r)$ is oscillatory, as in Eq. \eqref{Eq:AsymptoticDecayh(r)Osc}, so that the integral in Eq. \eqref{Eq:IsothermalCompressibility} is negative. For mechanical stability, $\chi_T$ must always be positive. Such observations suggest a crossover from monotonic to damped oscillatory decay of $h(r)$ might be reflected in the behavior of the thermodynamic quantity $\chi_T/\chi_T^\text{id}$ across the phase diagram.

We explore this possibility, focusing on the excess, over ideal, compressibility defined by
\begin{equation}
	\chi_T \equiv \chi_T^\text{id} + \chi_T^\text{ex}\,,
\end{equation}
and argue that the line in the phase diagram where $\chi_T^\text{ex} = 0$ should lie near the FW line for simple liquids.
Recall that the relative location of the poles is controlled by the form of $c(r)$ (see discussion in Sec. \ref{Sec:ParticleForcesAndDecay}). A pure imaginary pole, $q = i \alpha_0^\text{mon}$, is determined by Eq. \eqref{Eq:DefPolesRealSpace_1} with $\alpha_1 = 0:$
\begin{equation} 
		1 = 4 \pi\rho_b\int_{0}^{\infty}\text{d}r\,r^2c(r)\frac{\sinh(\alpha_0^\text{mon}\, r)}{\alpha_0^\text{mon}\, r}\,. \label{Eq:DefImaginaryPole}
\end{equation}
Suppose first we are at a state point where repulsion dominates so that $c(r)$ is dominated by its negative core contribution $r < \sigma$. Then it is likely that the smallest solution $\alpha_0^\text{mon}$ will be greater than the imaginary parts $\alpha_0^\text{osc}$ of the complex poles obtained from solving Eqs. \eqref{Eq:DefPolesRealSpace_1} and \eqref{Eq:DefPolesRealSpace_2} and therefore the ultimate decay of $h(r)$ will be damped oscillatory. This follows since the right-hand side of Eq. \eqref{Eq:DefImaginaryPole} must be equal to one and thus the integral is positive definite. But this can be achieved only for large values of $\alpha_0^\text{mon}$ if $c(r)$ has only small positive contributions arising from weak attraction. In order to obtain a leading imaginary pole, the interparticle attraction must be sufficiently strong to counterbalance the negative core contributions in $c(r)$. 

In this context it is useful to consider a near-critical state point where $\alpha_0^\text{mon}$ is small. Expanding $\sinh(\alpha_0^\text{mon} r)$ in Eq. \eqref{Eq:DefImaginaryPole}  to second order one finds
\begin{equation} \label{Eq:(ii)}
	\left(\alpha_0^\text{mon}\right)^2 \frac{2\pi}{3}\rho_b \int_{0}^{\infty}\text{d}r\,r^4\,c(r) = 1 - C\,, 
\end{equation}
with 
\begin{equation} \label{Eq:Def_C}
	C = 4\pi\rho_b\int_{0}^{\infty}\text{d}r\,r^2 c(r) = \rho_b\widehat{c}(0)\,.
\end{equation}
Recalling that the static structure factor $S(k)$ is given by \cite{HansenMcDonald2013}
\begin{equation} \label{Eq:DefS(k)}
	S(k) = \frac{1}{1 - \rho_b \widehat{c}(k)}\,,
\end{equation}
it follows that $1 - C = S(0)^{-1}$, which must be positive for the fluid to be stable. Thus, Eq. \eqref{Eq:(ii)} has real solutions, which we denote $\alpha_{OZ}$, provided the second moment of $c(r)$ is positive. This requires $c(r)$ to be sufficiently positive at large $r$, i.e. there must be sufficient attraction. Note that the solution of Eq. \eqref{Eq:(ii)} then yields the second moment or Ornstein-Zernike (OZ) correlation length:
\begin{equation} \label{Def:xi_OZ}
	\xi_{OZ}^2 \equiv \alpha_{OZ}^{-2} = R^2 S(0)\,,
\end{equation}
where $R^2 = \frac{2\pi}{3} \rho_b \int_{0}^{\infty}\text{d}r\,r^4 c(r)$ defines the short-ranged correlation length or Debye persistence length \cite{HansenMcDonald2013} \red{and expanding $\widehat{c}(k)$ to $\mathcal{O}(k^2)$ in Eq. \eqref{Eq:DefS(k)} yields the celebrated OZ formula for the structure factor: $S(k)=S(0)/(1 + \xi_{OZ}^2 k^2)$, $k\rightarrow 0$.} By considering the Taylor expansion of $x\sinh(x)$ it is easy to show $\alpha_0^\text{mon} < \alpha_{OZ}$, i.e. the true \cite{Footenote} correlation length $\xi$, obtained from Eq. \eqref{Eq:DefImaginaryPole}, is larger than the OZ one, obtained from Eq. \eqref{Eq:(ii)}: $\xi>\xi_{OZ}$.

These considerations point to the importance of having a positive second moment of $c(r)$ in order to obtain monotonic decay of $h(r)$ with a long correlation length. By contrast, for a model fluid that exhibits purely repulsive interactions, such as hard-spheres, $c(r)$ is negative apart from a very weak, rapidly decaying tail outside the hard core, so that both the second and first moments of $c(r)$ are negative. Then the only poles are complex, $\alpha_1 > 0$, and are determined by solving Eqs. \eqref{Eq:DefPolesRealSpace_1} and \eqref{Eq:DefPolesRealSpace_2}.

For a model fluid that exhibits, both repulsive and attractive interactions, it is clear that the first moment Eq. \eqref{Eq:Def_C} at a given state point provides a measure of the competition between repulsive and attractive contributions to $c(r)$. $C$ will be positive when attraction dominates but negative when repulsion dominates. From plots such as those in Fig. \ref{Fig:cr_MC} for $c(r)$ in the SW fluid one can surmise that $C$ changes from positive values at low packing fractions $\eta$ to negative values at large $\eta$. And we know from Fig. \ref{Fig:gr_MC} that FW crossover occurs at an intermediate $\eta$. We conjecture that generally the change of sign of $C$ should reflect the change from monotonic to damped oscillatory decay of $h(r)$. Since the criterion $C = 0$ corresponds to $S(0) = 1$, see Eq. \eqref{Eq:DefS(k)}, this implies FW crossover should occur when the $k = 0$ limit of the structure factor is near the ideal gas value. Using the compressibility sum rule \cite{HansenMcDonald2013} $S(0) = \rho_b k_B T \chi_T$, identical to \eqref{Eq:IsothermalCompressibility}, it follows that the line defined by $C = 0$ corresponds to the line of vanishing excess compressibility: $\chi_T^\text{ex} = 0$.
From the arguments above it is clear that this line cannot be identical to the FW line, defined by equality of asymptotic decay lengths, i.e. $(\alpha_0^\text{osc})^{-1} = (\alpha_0^\text{mon})^{-1}$ but we conjecture the lines will be close.

In the next section, we examine how close these two lines are for several model fluids. To this end, we employ classical density functional theory, which is a powerful framework to describe structure and thermodynamics on equal footing. We investigate the square-well fluid in detail, but consider also the hard-core Yukawa fluid, the sticky hard-sphere fluid, and the Asakura-Oosawa model.

\section{Results for model fluids} \label{Sec:Results}

\subsection{The square-well fluid} \label{SubSec:SW}

We consider first the square-well (SW) fluid which is the crudest model system for describing simple fluids such as argon. The pair interaction potential $\phi(r)$ is given by 
\begin{align}
	\phi(r) 
	 &= 
	\begin{cases}
	\infty~~~~~;~r < \sigma \\
	-\varepsilon~~~~;~\sigma < r < \lambda\sigma \\
	0~~~~~~;~r > \lambda\sigma\,,\\
	\end{cases}
\end{align}
where $\varepsilon$ is the strength of the attraction which acts in the range $\sigma < r < \lambda\sigma$. The hard-core diameter is $\sigma$. 
We choose to employ the powerful framework of classical density functional theory \cite{Evans1979}, which is based on the theorem that the structure and thermodynamics of a fluid can be obtained by minimizing the grand-potential functional $\Omega[\rho]$ of the one-body density $\rho(\mathbf{r})$,
 \begin{equation} \label{Eq:grandPotentialFunctional}
 	\Omega[\rho] = F[\rho] + \int\text{d}\mathbf{r}\,\rho(\mathbf{r}) \left(V_\text{ext}(\mathbf{r}) - \mu\right)\,.
 \end{equation}
 The equilibrium density profile $\rho_\text{eq}(\mathbf{r})$ satisfies
 \begin{equation} \label{Eq:EulerLagrange}
 	0 = \left.\frac{\delta \Omega[\rho]}{\delta\rho(\mathbf{r})}\right|_{\rho=\rho_\text{eq}}\,.
 \end{equation} 
 Here, $V_\text{ext}(\mathbf{r})$ denotes an arbitrary external potential acting on the fluid particles, $\mu$ is the chemical potential (of the corresponding particle reservoir) and $F[\rho] = F_\text{id}[\rho] + F_\text{ex}[\rho]$ is the intrinsic Helmholtz free-energy functional, split into an exactly known ideal-gas part
 \begin{equation}
 	\beta F_\text{id}[\rho] = \int\text{d}\mathbf{r} \rho(\mathbf{r}) \left[\ln\left(\rho(\mathbf{r})\Lambda^3\right) - 1\right]\,,
 \end{equation}
 where $\Lambda$ is the thermal wavelength of the particles, and an excess part $F_\text{ex}[\rho]$ which contains all information about the particle interactions. We describe the model fluid via the standard mean-field approach \cite{HansenMcDonald2013, ArcherEvans2017}, i.e.
 \begin{equation} \label{Eq:FreeEnergyFunctional_SW}
 	F_\text{ex}[\rho] = F_\text{\tiny HS}[\rho] + \frac{1}{2}\iint \text{d}\mathbf{r}\text{d}\mathbf{r}' \rho(\mathbf{r})\rho(\mathbf{r}')\phi_\text{att}^\text{sw}(|\mathbf{r}-\mathbf{r}'|)\,,
 \end{equation}
 where $\phi_\text{att}^\text{sw}(r)$ denotes the attractive portion of the SW pair potential. Of course, there is flexibility in defining this. Here we split the total pair potential $\phi(r) = \phi_\text{\tiny HS}(r) + \phi_\text{att}^\text{sw}(r)$ into the hard-sphere contribution and an attractive tail given by:
 \begin{align}
 	&\phi_\text{\tiny HS}(r) = \begin{cases}
 	\infty ~~;~~r < \sigma \\
 	0~~~~;~~r > \sigma\,,\\
 	\end{cases}
 	;
 	&\phi_\text{att}^\text{sw}(r) = \begin{cases}
 	-\varepsilon~~;~~r < \lambda\sigma\\
 	0~~~~;~~r > \lambda\sigma\,.
 	\end{cases}
 \end{align}
 Extending the attraction to inside the core compensates underestimation of correlations.
 For the hard-sphere part of the excess free-energy functional, $F_\text{\tiny HS}[\rho]$, we employ the  White-Bear Mark II functional \cite{HansenGoos2006WB2} which is a more accurate version of Rosenfeld's fundamental measure theory (FMT) \cite{Rosenfeld1989}.
 For bulk (uniform) fluids, with a constant bulk density $\rho_b$, Eq. \eqref{Eq:FreeEnergyFunctional_SW} yields to the excess free energy density
 \begin{equation}
 	\beta f_\text{ex} = \frac{\beta F_\text{ex}[\rho_b]}{V} = \rho_b \frac{4\eta-3\eta^2}{(1-\eta)^2} + \frac{1}{2} \rho_b^2\, \beta \widehat{\phi}_\text{sw}(0)\,,
 \end{equation}
 where $\eta = \pi\sigma^3\rho_b/6$ is the fluid packing fraction, and $\widehat{\phi}_\text{att}^\text{sw}(0) = 4\pi\int_{0}^{\infty}\text{d}r\,r^2 \phi_\text{att}^\text{sw}(r)$ is the $k = 0$ limit of 
 \begin{equation}
 	\widehat{\phi}_\text{att}^\text{sw}(k) = -\frac{4\pi \varepsilon}{k^3}\left[\sin(\lambda\sigma k) - \lambda\sigma k\,\cos(\lambda\sigma k)\right]\,,
 \end{equation}
 the three-dimensional Fourier transform of the pair potential $\phi_\text{att}^\text{sw}(r)$. The pressure of the SW fluid is then given by the generalized van der Waals form
 \begin{equation} \label{Eq:EquationOfState_SW}
 	\beta p = \rho_b \frac{1+\eta+\eta^2-\eta^3}{(1-\eta)^3} - 4 \beta\varepsilon\rho_b\eta\lambda^3\,,
 \end{equation}
 where the first term is the accurate Carnahan-Starling (CS) reduced pressure of hard-spheres.
 The isothermal compressibility $\chi_T$ is easily calculated using Eq. \eqref{Eq:IsothermalCompressibility}. The condition $\chi_T^\text{ex} = 0$ is equivalent to $\rho_b k_B T \chi_T = 1$, which is identical to 
 \begin{align} \label{Eq:StopperLine_SW}
 	\frac{\partial \beta p}{\partial \rho_b} = 1~~~
 	\Leftrightarrow~~~ \beta\varepsilon\lambda^3 = \frac{1 - \frac{\eta}{4}}{(1-\eta)^4}\,.
 \end{align}
 Thus, for the SW model treated in MF DFT, the line where the isothermal compressibility takes its ideal gas value is described by a simple formula. The form of Eq. \eqref{Eq:StopperLine_SW} is a direct result of the simple generalized van der Waals equation of state \eqref{Eq:EquationOfState_SW}, where the attractive contribution to the pressure is proportional to $\rho_b^2$. If we consider a general form $\phi_\text{att}(r)$ for the attractive potential in Eq. \eqref{Eq:FreeEnergyFunctional_SW} the condition $\chi_T^\text{ex} = 0$ reduces to
 \begin{equation} \label{Eq:StopperLine_MF}
 	-\beta\widehat{\phi}_\text{att}(0)\sigma^{-3} = \frac{4\pi}{3} \frac{1-\frac{\eta}{4}}{(1-\eta)^4}\,,
 \end{equation} 
 with $\widehat{\phi}_\text{att}(0) = 4\pi\int_{0}^{\infty}\text{d}r\,r^2\phi_\text{att}(r)$. In Appendix \ref{Sec:Appendix_A} we show that Eq. \eqref{Eq:StopperLine_MF} leads to a law of corresponding states for the line $\chi_T^\text{ex}  = 0$.
 
It is clear that determining this line is much simpler than calculating the FW line where the leading poles of $\widehat{h}(k)$ must be determined. The former is determined by a thermodynamic criterion, i.e. only the bulk free energy density is required. By contrast, and as discussed in Sec. \ref{Sec:ParticleForcesAndDecay}, the FW line is determined by the bulk correlation function $c(r)$ that must be provided by the underlying microscopic theory. Within the framework of DFT, $c(r)$ can be obtained by functional differentiation of the (approximate) excess free-energy functional $F_\text{ex}[\rho]$:
 \begin{equation}
 c(\eta\,;\,r) = \left.-\frac{\delta^2 \beta F_\text{ex}[\rho]}{\delta\rho(\mathbf{r}) \delta\rho(\mathbf{r}')}\right|_{\rho_b}\,.
 \end{equation}
For the MF DFT \eqref{Eq:FreeEnergyFunctional_SW}, $c(r)$ has a relative simple analytical form, 
\begin{equation} \label{Eq:DirectCorrelation_SW}
	c(\eta\,;\,r) = c_\text{\tiny HS}(\eta\,;\,r) -\beta\phi_\text{att}^\text{sw}(r)\,,
\end{equation}
where $c_\text{\tiny HS}(r)$, obtained from White-Bear Mark II, is a third-order polynomial of $r$ with range $\sigma$, depending on the packing fraction $\eta$ and the hard-sphere diameter $\sigma$. Note that Eq. \eqref{Eq:DirectCorrelation_SW} defines the Random Phase Approximation (RPA); see Refs. \onlinecite{HansenMcDonald2013, ArcherEvans2017}. Results for $c(r)$ from this MF DFT are shown in Fig. \ref{Fig:cr_MC} (a).

\begin{figure}[t] 
	\centering
	\includegraphics[width = 8cm]{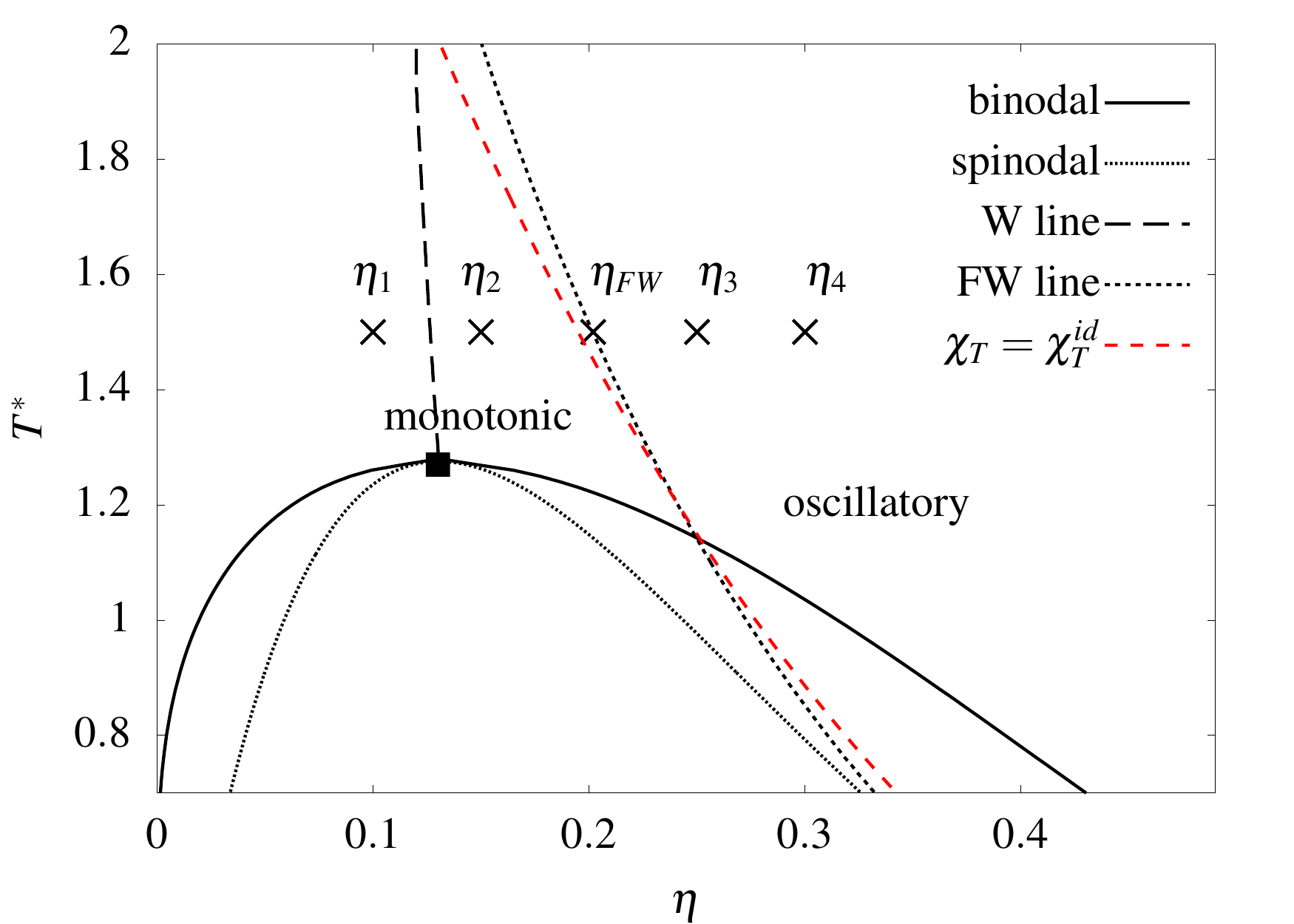}
	\caption{The phase diagram of the square-well fluid \textcolor{black}{with $\lambda=1.5$} treated within MF DFT fashion in the $T^*$-$\eta$ plane. The solid line shows the binodal, and the fine dotted line below the binodal is the spinodal. The long-dashed line terminating at the critical point black square $(\eta_c, T_c^*) = (0.13, 1.27)$ is the Widom line, i.e. the line of local maximal correlation length $\xi = \alpha_0^{-1}$. The short dashed-line is the Fisher-Widom line separating regions of monotonic and oscillatory decay, and the medium-dashed line is where $\chi_T = \chi_T^\text{id}$. The crosses labeled $\eta_1$, $\eta_2$, $\eta_3$, $\eta_4$, and $\eta_\text{FW}$ denote state points considered in this work.} \label{Fig:phase_diag_SW}
\end{figure}

In Fig. \ref{Fig:phase_diag_SW} we plot the phase diagram for the SW fluid \textcolor{black}{with $\lambda=1.5$} in the $T^*$-$\eta$ plane where $T^* = k_B T/\varepsilon$, also recently considered by Roth in Ref. \onlinecite{Roth2018}. The critical point is located at $(\eta_c, T^*_c) = (0.13, 1.27)$. The solid line shows the binodal, where the gas and liquid coexist. The fine dotted line below the binodal is the spinodal, which is most easily determined by searching for solutions where the isothermal compressibility diverges. 
The Fisher-Widom line is the short-dashed line, which is bounded by the spinodal at low $T^*$. This follows since the spinodal can also be defined as the solutions of Eq. \eqref{Eq:DefPolesRealSpace_1}  with $\alpha_0 = 0$, corresponding to $1 - \rho_b \widehat{c}(0) = 0$. On the low density side of the FW line, correlations decay purely monotonically (where $\alpha_1 = 0$), whereas on the high density side the decay is damped oscillatory. The FW line is determined numerically by searching for solutions of $1 - \rho_b \widehat{c}(k) = 0$ where complex poles and imaginary poles have the same (smallest) imaginary part, $\alpha_0^\text{mon} = \alpha_0^\text{osc}$. 
\textcolor{black}{The long-dashed line in Fig.~\ref{Fig:phase_diag_SW} emanating almost vertically from the critical point is the Widom (W) line, which we defined as the line of a local maximum of the true \cite{Footenote} correlation length $\xi$ (see Introduction \ref{Sec:Introduction}), corresponding to a minimum of $\alpha_0$. At higher temperatures, the W line approaches the FW line}.

 Strikingly, the line where $\chi_T^\text{ex}$  vanishes (medium dashed-line) is rather close to the FW line. At higher temperatures the deviations become larger, but in the important region approaching coexistence these are fairly small. In particular, there is no systematic offset between the FW line and the line where $\chi_T^\text{ex} = 0$. This means that at state points where $c(r)$ has sufficient positive (attractive) contributions such that the integral in Eq. \eqref{Eq:Def_C} vanishes, we are likely to find FW crossover in close proximity to the $\chi_T^\text{ex} = 0$ criterion.
\begin{figure}[t] 
	\centering
	\includegraphics[width = 8cm]{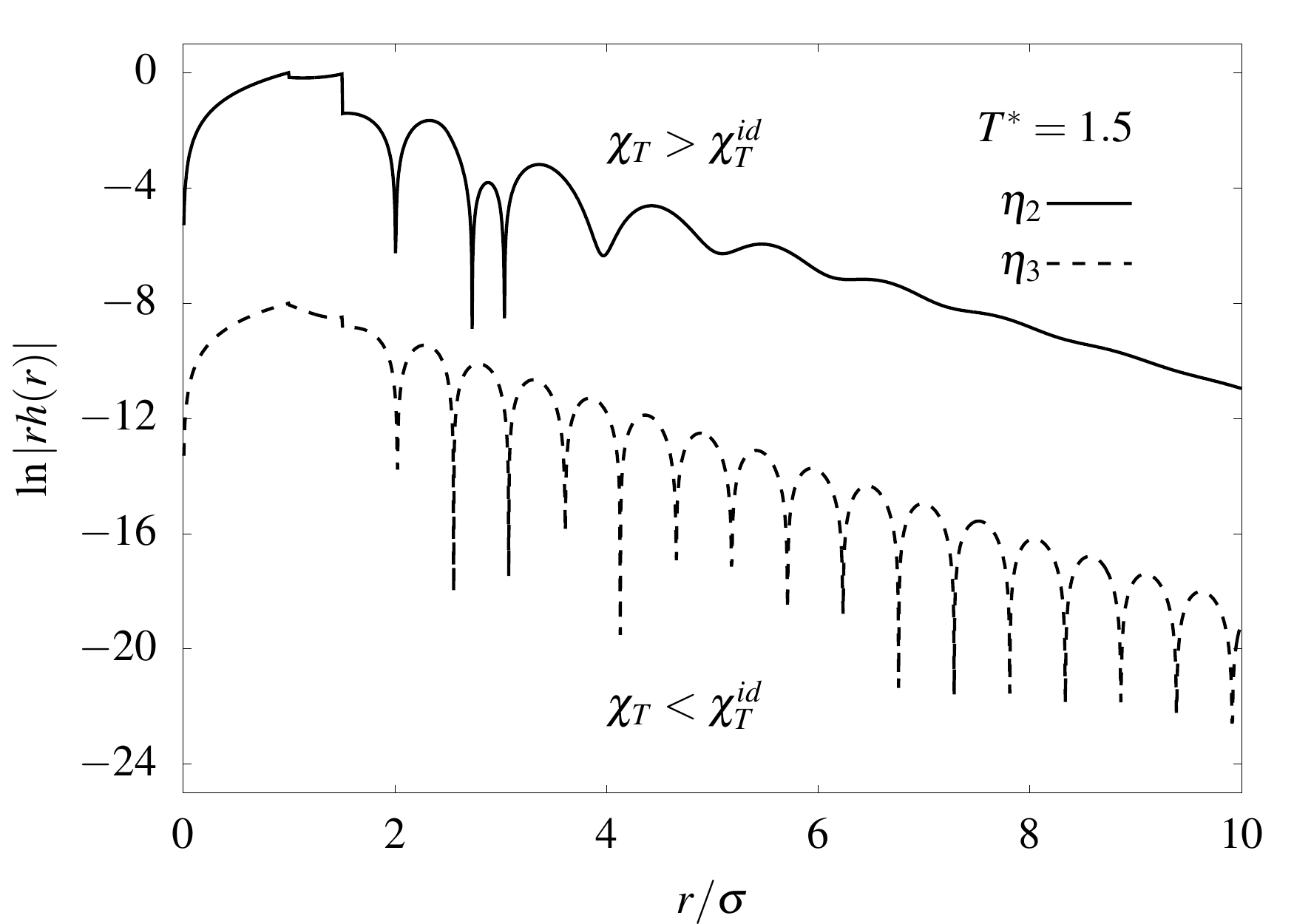}
	\caption{\textcolor{black}{Asymptotic decay of $h(r)$ for the SW fluid, as obtained from MF DFT by minimizing the grand-potential functional in the presence of a test particle. The curves are shifted vertically for clarity. The state points, both at temperature $T^* = 1.5$, are located on the monotonic side of the FW line ($\eta_2 = 0.15$; upper curve), and the oscillatory side of the FW line ($\eta_3 = 0.25$; bottom curve).} } \label{Fig:gr_MF_DFT}
\end{figure} 
\begin{figure}[t] 
	\centering
	\includegraphics[width = 8cm]{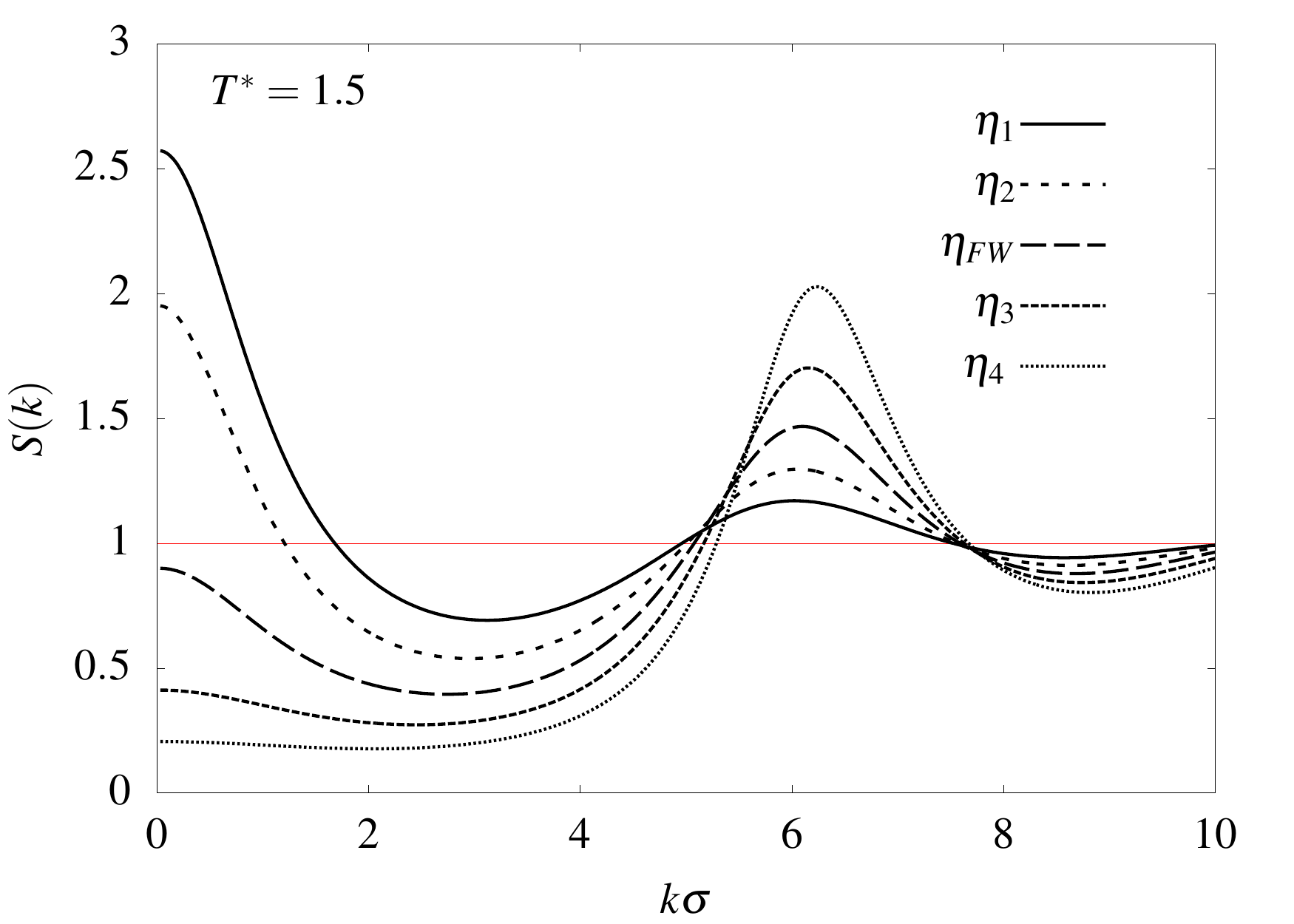}
	\caption{Static structure factor $S(k) = [1 - \rho_b\widehat{c}(k)]^{-1}$ of the SW fluid within MF DFT for the state points (crosses) shown in Fig. \ref{Fig:phase_diag_SW} with fixed temperature $T^* = 1.5$.} \label{Fig:Sk_MF_DFT}
\end{figure} 
Fig. \ref{Fig:phase_diag_SW} indicates nicely how the FW line is bounded by the liquid spinodal \cite{Evans1993}. It is obvious from Sec. \ref{Sec:ParticleForcesAndCompr} that the line $\chi_T=\chi_T^\text{id}$ must lie outside the spinodal, i.e. at larger $\eta$, and we observe it also lies outside the FW line at low $T^*$.

In Fig. \ref{Fig:gr_MF_DFT} we plot the asymptotic decay of $h(r)$ as obtained from numerically minimizing the grand-potential functional in presence of a test particle -- the Percus' test-particle procedure \cite{HansenMcDonald2013}. This corresponds to fixing a particle at the origin so that it exerts an external potential $V_\text{ext}(r) = \phi(r)$, the pair interaction potential, on other particles. Then the one-body density profile satisfies $\rho(\mathbf{r}) = \rho(r) = \rho_b g(r)$, with $g(r) = h(r) + 1$. It is important to recognize that the inverse decay length and the wavelength that can be extracted from the decay of (test particle) plots such as Fig. \ref{Fig:gr_MF_DFT} correspond precisely to $\alpha_0$ and $2\pi/\alpha_1$ determined by calculating the poles of $\widehat{h}(k)$ or zeros of $1 - \rho_b \widehat{c}(k)$ with $\widehat{c}(k)$ from Eq. \eqref{Eq:DirectCorrelation_SW}, usually termed the OZ route. The equivalence between the test particle and OZ routes for asymptotic decay is discussed in a recent paper \cite{WaltersEvans2018}, \red{and is based on} linear response arguments \cite{EvansCarvalho1996}.
We consider two state points, marked as crosses in Fig. \ref{Fig:phase_diag_SW}, at constant temperature $T^* = 1.5$: One is located on the monotonic side of the FW line at volume fraction  $\eta_1 = 0.15$, and the other on the oscillatory side of the FW line at $\eta_2 = 0.25$.
 As found in the simulations (cf. Fig. \ref{Fig:gr_MC}), we observe a crossover from monotonic to damped oscillatory decay as $\eta$ is increased. Calculating $h(r)$ precisely at the point where FW crossover occurs, i.e. where both types of poles (complex and imaginary) contribute to the decay for $r\rightarrow \infty$, we find that $h(r)$ decays overall in a damped oscillatory fashion. At FW crossover $r\,h(r)$ decays as
 \begin{align}
 	rh(r)&\sim A_\text{mon} \exp(-\alpha_0^\text{mon}\, r) \notag\\ &+ A_\text{osc}\exp(-\alpha_0^\text{osc}\, r) \cos(\alpha_1\, r - \theta)\,,~~r\rightarrow\infty\,,
\end{align}
 where \textcolor{black}{$\alpha_0^\text{mon}=\alpha_0^\text{osc}$}, and $A_\text{mon}$ and $A_\text{osc}$ are amplitudes. \textcolor{black}{Note that on the FW line one still finds damped oscillatory (non-monotonic) decay of $rh(r)$ as $r\rightarrow\infty$, regardless of the (non-zero) values of $A_\text{mon}$ and $A_\text{osc}$. If $A_\text{osc}>A_\text{mon}$ then $rh(r)$ has zeros, which is not the case if $A_\text{osc}<A_\text{mon}$}. 
 At this state point, the isothermal compressibility is close to its ideal-gas value; see Fig. \ref{Fig:phase_diag_SW}. This is also demonstrated in Fig. \ref{Fig:Sk_MF_DFT} where the static structure factor $S(k)$ is plotted for the five state points shown in Fig. \ref{Fig:phase_diag_SW}. We use the OZ route, i.e. $S(k)$ is given by Eq. \eqref{Eq:DefS(k)}. Within this framework, $S(k)$ can be computed analytically. On the monotonic side of the FW line (packings $\eta_1$ and $\eta_2$) the value of $S(k = 0) = \chi_T/\chi_T^\text{id}$ is substantially greater than unity. At high packing fractions ($\eta_3$ and $\eta_4$), $S(k = 0) < 1$, and decreases rapidly with increasing $\eta$, reflecting the rapid decrease of $\chi_T$. For $\eta = \eta_{FW}$ we find $S(k = 0)$ is close to unity. Note that the position and the height of the maximum of $S(k)$ increases smoothly with increasing $\eta$.

 \subsection{Colloid-polymer mixtures: the AO model} \label{SubSec:AO}
 
 In Sec. \ref{SubSec:SW} we considered a one-component SW fluid, whose physics is described by a rather simple generalized van der Waals theory. Competition between repulsive and (effective) attractive interactions can also occur in binary (or multiple component) systems where the bare interactions are purely repulsive. This so-called depletion effect is a purely entropy-driven mechanism leading to attraction between particles of one species, usually the bigger, induced by the other smaller species \cite{Lekkerkerker2011}. A prominent example is the model colloid-polymer mixture described by Asakura and Oosawa \cite{AasakuraOosawa1954, AasakuraOosawa1958} and later by Vrij \cite{Vrij1976}. The colloids ($C$) are considered as hard-spheres with a diameter $\sigma_C$. The polymers ($P$) are represented by spheres with diameter $\sigma_P$ which also interact with the colloids as hard-spheres but do not interact with other polymers. This type of interaction yields a region around each colloid which cannot be entered by the polymers, termed the depletion zone; it has a radius of $(\sigma_C + \sigma_P)/2$ around each colloid. When two colloids are sufficiently close together so that their depletion zones overlap, there is an increase of accessible volume for the polymers. As a result, the total entropy of the system increases, leading to an effective attraction between the colloids. At sufficiently large size ratios $q = \sigma_P/\sigma_C \gtrsim 0.3$ there is stable, w.r.t. crystallization, colloidal fluid-fluid demixing into colloid-rich and colloid-poor phases \cite{Gast1983, LekkerkerkerPoon1992}, the colloidal equivalent of gaseous and liquid phases in atomic fluids.
 The thermodynamics and microscopic structure of the binary AO model is described successfully over a wide range of size ratios $q$, by a non-mean-field density functional \cite{Schmidt2000} constructed within the framework of Rosenfeld's FMT. 
 The starting point is the FMT excess free energy functional for hard-spheres which has the general form
 \begin{equation}
 	\beta F_\text{ex}[\{\rho_i\}] = \int\text{d}\mathbf{r}\,\Phi(\{n_\nu^i(\mathbf{r})\})\,,
 \end{equation}
 where $\Phi(\{n_\nu^i\})$ is a function of a set of weighted densities $n_\nu^i(\mathbf{r}) = \int\text{d}\mathbf{r}'\,\rho_i(\mathbf{r}')\omega_\nu^i(\mathbf{r}-\mathbf{r}')$. The $\omega_\nu^i(\mathbf{r})$ are weight functions \cite{Rosenfeld1989} characterizing the geometry of a sphere of species $i$ with radius $\sigma_i/2$. The functional for the AO model is described by the following reduced free energy density:
 \begin{equation} \label{Eq:AO_Functional}
 	\Phi_\text{AO}(\{n_\nu^C, n_\nu^P\}) = \Phi(\{n_\nu^C \}) + \sum_{\nu}\frac{\partial \Phi(\{n_\gamma^C\})}{\partial n_\nu^C}\,n_\nu^P\,,
 \end{equation}
 i.e. the FMT functional (in this paper the White-Bear Mark II \cite{HansenGoos2006WB2}) for a binary hard-sphere fluid is linearized \cite{Schmidt2000, SchmidtBrader2002} with respect to the weighted density of one component (the polymers $n_\nu^P$). For the uniform fluid we obtain from Eq. \eqref{Eq:AO_Functional} the excess semi-grand potential density
 \begin{equation} \label{Eq:FreeEnergyAO}
 	\beta\omega_\text{ex}^\text{AO}(\rho_C,\,\rho_P^r) = \rho_C \frac{4\eta_C - 3\eta_C^2}{(1-\eta_C)^2} - \rho_P^r\, \alpha(\eta_C)\,,
 \end{equation}
 where $\rho_P^r$ is the number density of the polymers in the reservoir and $\alpha(\eta_C)$ is the free-volume fraction, i.e. the volume that is accessible to the polymers in the system due to the presence of a \red{packing} fraction $\eta_C$ of colloids. For the AO model described within FMT, $\alpha(\eta_C)$ can be calculated from \cite{OversteegenRoth2005}
 \begin{equation}
 	\alpha(\eta_C) = \exp\left[-\sum_{\nu} \frac{\partial \Phi(\{n_\gamma^C \})}{\partial n_\nu^C} \frac{n_\nu^P}{\rho_P}\right]\,,
 \end{equation}
 where the weighted densities are evaluated for $\rho_C(\mathbf{r}) = \rho_C$ and $\rho_P(\mathbf{r}) = \rho_P$. For the line of vanishing excess isothermal compressibility $\chi_T^\text{ex} = 0$ we find
 \begin{equation}
 	\frac{\eta_P^r}{8q^3}\,\alpha''(\eta_C) = \frac{1-\frac{\eta_C}{4}}{(1-\eta_C)^4}\,,
 \end{equation}
 where $\eta_P^r = \pi \sigma_P^3 \rho_P^r/6$ is the polymer packing fraction of the reservoir and the prime denotes the derivative w.r.t. $\eta_C$.

 The bulk pair direct correlation functions $c_{ij}(r)$ with $i,\,j=C,\,P$ take the following form \cite{SchmidtBrader2002}
\begin{align}
	c_{CC}(\eta_C\,,\eta_P\,;\,r) &= c_\text{\tiny HS}(\eta_C\,;\,r) + \eta_P\, c_*(\eta_C\,;\,r)\,, \\
	c_{CP}(\eta_C\,,\eta_P\,;\,r) &= c_{CP}(\eta_C\,;\,r)\,,\\
	c_{PP}(\eta_C\,,\eta_P\,;\,r) &= 0\,,
\end{align}
where $c_\text{\tiny HS}(r)$ is the FMT solution for the pair direct correlation function of the one-component hard-sphere system. Note that in this work $c_\text{\tiny HS}(r)$ is different from Refs. \onlinecite{Schmidt2000, SchmidtBrader2002} where the less accurate PY result for $c_\text{\tiny HS}(r)$ was used. It is the linearity of the AO functional in the polymer density $\rho_P(\mathbf{r})$ that dictates that $c_{PP}(r)$ vanishes, $c_{CP}(r)$ depends only on the packing \red{fraction} of colloids $\eta_C$, and $c_{CC}(r)$ is linear in $\eta_P$. The function $c_*(r)$ describes the portion of the direct correlation between two colloids due to the presence of the polymers. As the asymptotic decay of $h_{ij}(r)$ in a binary mixture is governed by the zeros of a common denominator $D(k)$ for all $\widehat{h}_{ij}(k)$, the decay length and the wavelength of any oscillatory decay of $r h_{ij}(r)$, for $r\rightarrow\infty$, is independent of the species \cite{Evans1994}. For the AO model, $D(k)$ is given generally by:
\begin{equation}
	D(k) = [1 - \rho_C\, \widehat{c}_{CC}(k)][1-\rho_P \widehat{c}_{PP}(k)] - \rho_C\rho_P\,[\widehat{c}_{CP}(k)]^2\,,
\end{equation}
where $\rho_C$ and $\rho_P$ are the bulk densities of the colloids and polymers, respectively. $D(k)$ simplifies in the linearization since $\widehat{c}_{PP}(k) = 0$.
\begin{figure}[t] 
	\centering
	\includegraphics[width = 8cm]{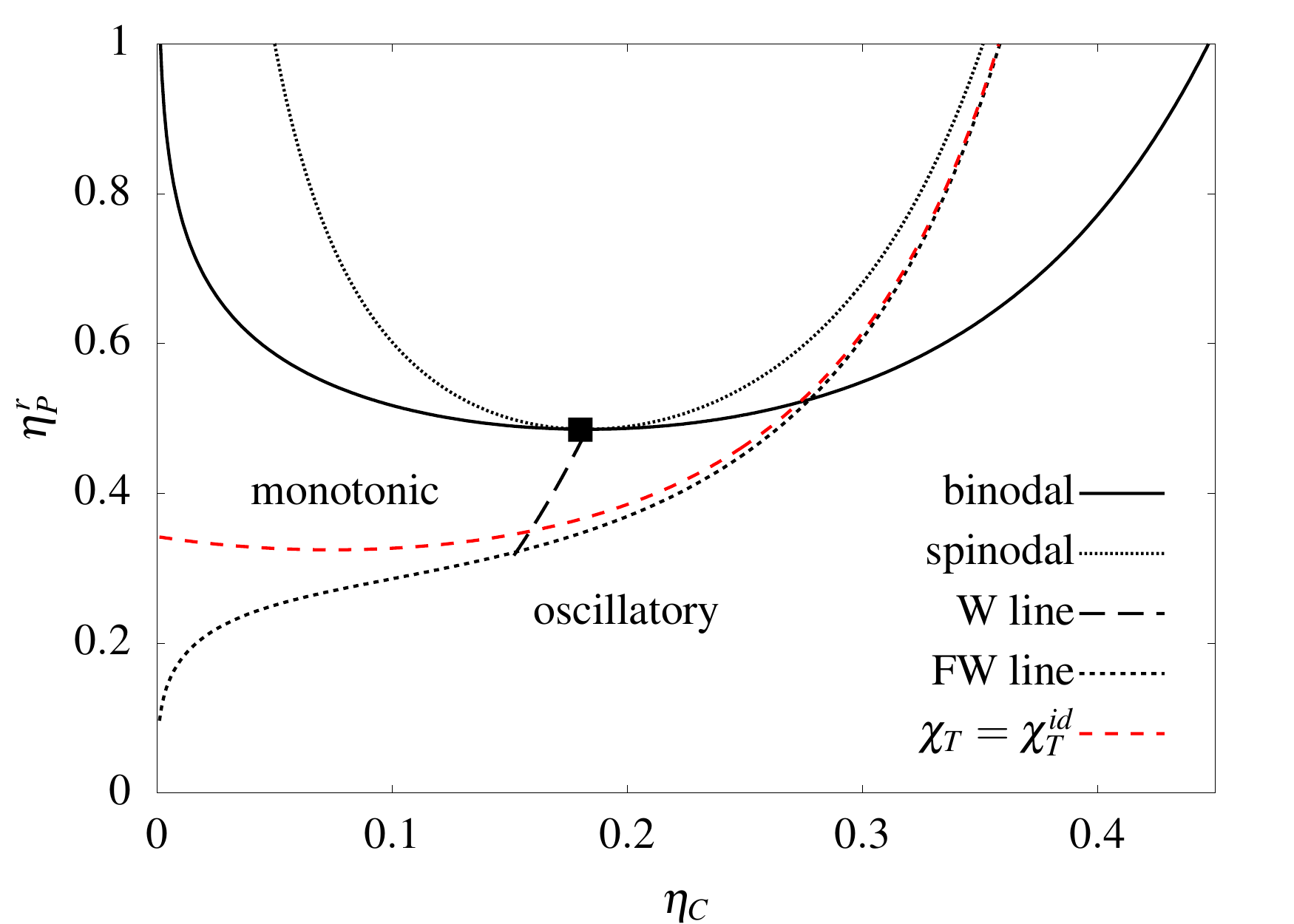}
	\caption{Phase diagram of the AO model for a size ratio $q = 0.6$ in the $\eta_P^r$-$\eta_C$ (polymer reservoir-colloid fraction) plane displaying the same curves as in Fig. \ref{Fig:phase_diag_SW} for the SW; the ordinate corresponds to inverse temperature. Note that the Widom line \red{terminates at the FW line -- see text}.} \label{Fig:phase_diag_AO}
\end{figure} 
In Fig. \ref{Fig:phase_diag_AO} we plot the phase diagram of the AO model for a size ratio $q = 0.6$ in the $\eta_P^r$-$\eta_C$ plane with the accompanying FW, W and $\chi_T^\text{ex} = 0$ lines. It is well-known that the polymer reservoir density $\eta_P^r$ plays the role of an inverse temperature \cite{Gast1983, LekkerkerkerPoon1992}. The critical point (black square) is located at $(\eta_C, \eta_P^r) = (0.18, 0.49)$. It is very close to that shown in Fig. 1 of Ref. \onlinecite{SchmidtBrader2002} where the FMT is that of Rosenfeld. Similar to what we have observed in Sec. \ref{SubSec:SW}, the line where $\chi_T^\text{ex} = 0$ is very close to the FW line for colloidal packing fractions $\eta_C$ larger than the critical value and within the high-density metastable region inside the binodal these two lines are nearly indistinguishable. Moreover this observation is valid for the other size ratio $q$ that we investigated. In contrast to the MF DFT result for SW in Fig. \ref{Fig:phase_diag_SW}, we do not observe a clear intersection between the line of vanishing excess isothermal compressibility (red-dashed) and the FW line (dotted). It is important to note the behavior of the FW (dotted) and $\chi_T = \chi_T^\text{id}$ (dashed) lines as $\eta_C\rightarrow 0$. The former decreases steeply, indicating crossover shifts to very low values of the polymer reservoir volume fraction. This is completely consistent with what is found for simple liquids. The results in Fig. 7 of Ref. \onlinecite{Dijkstra2000} for a truncated Lennard-Jones fluid, treated within the RPA, show the FW line approaching the temperature axis asymptotically as the density of the fluid goes to zero, $\rho_b\rightarrow 0$. The same behavior is found by Roth for the SW fluid \cite{Roth2018} and in the present work; this is not visible on the scale of the temperature axis in Fig. \ref{Fig:phase_diag_SW}, shown in Sec.~\ref{SubSec:SW}. In the limit $T\rightarrow\infty$, correlations must be hard-sphere like \cite{Dijkstra2000}. For the AO model $\eta_P^r$ plays the role of an inverse temperature and thus one expects the asymptotic approach of the FW line to $\eta_P^r = 0$ as $\eta_C\rightarrow 0$ shown in Fig. \ref{Fig:phase_diag_AO}. In contrast, the $\chi_T=\chi_T^\text{id}$ line exhibits no special feature as $\eta_C\rightarrow 0$.

 \red{The Widom line (W) emerging from the critical point, displays a near linear behavior with decreasing $\eta_C$. This behavior is equivalent to that in Fig. \ref{Fig:phase_diag_SW} for the SW where the line of maximal correlation length shifts to smaller $\eta_C$ as $T$ increases. In Fig. \ref{Fig:phase_diag_AO} we show the W line terminating at the FW line. For a given supercritical value of $\eta_P^r$ the pure imaginary (monotonic) pole $\alpha_0^\text{mon}$ exhibits a minimum as a function of the colloid packing fraction $\eta_C$ and the corresponding inverse defines a (local) maximum decay length for correlations. Provided $\alpha_0^\text{mon}<\alpha_0^\text{osc}$ then we can identify a maximum of the true correlation length: $\xi\equiv(\alpha_0^\text{mon})^{-1}$. This constitutes the \textit{slowest decay} length at the given $\eta_P^r$. In contrast, $\alpha_0^\text{osc}$ decays monotonically with increasing $\eta_C$. It follows that if $\alpha_0^\text{mon}>\alpha_0^\text{osc}$, i.e. on the oscillatory (low $\eta_P^r$) side of the FW line, the slowest decay length $(\alpha_0^\text{osc})^{-1}$ does not exhibit a minimum as a function of $\eta_C$. This is why we show the W line existing in only the monotonic region of the phase diagram, and terminating at the FW line. The same scenario applies for the SW at supercritical temperatures \cite{Roth2018}: $\alpha_0^\text{osc}$ decays monotonically with $\eta$ so that no maximum of the slowest decay length occurs for states in the oscillatory region of the $T-\eta$ phase diagram. We return to the importance of the FW line in bounding the W line in Sec. \ref{Sec:Conclusion}}. 

\subsection{Sticky hard-sphere fluid} \label{SubSec:SHS}

Introduced by Baxter \cite{Baxter1968} in 1968, the sticky hard-sphere (SHS) model is defined by a pair potential

 \begin{equation}
  \beta \phi_{\text{SHS}}(r) = 
     \begin{cases}
        \infty & r < \sigma' \\
        \ln\left(12\tau \frac{\sigma-\sigma'}{\sigma} \right) & \sigma' < r < \sigma \\
        0 & r>\sigma \, ,
     \end{cases}
\end{equation}
where the limit $\sigma' \to \sigma$ is taken. The potential $\phi_{\text{SHS}}$ corresponds to a hard-sphere repulsion with an attractive interaction present when particles are in contact (``sticky'' interaction). The model is constructed such that the second virial coefficient of the system remains finite in the sticky limit, and the model has been applied widely to describe particles with very short ranged-attraction, including colloidal suspensions \cite{Buzzaccaro2007} and micellar systems \cite{Mallamace2000}.
Baxter obtained an analytical solution for $c(r)$ for the SHS fluid within the PY closure of the OZ equation \cite{Baxter1968}. In the regime of strong attraction, i.e.\ small $\tau$, Baxter's solution for $c(r)$ becomes unphysical at intermediate packing fractions. However, it does feature a critical ``temperature'' $\tau_c = \frac{1}{3} - \frac{\sqrt{2}}{6} \approx 0.0976$ and for $\tau>\tau_c$ the result for $c(r)$ is always physical. The agreement between the compressibility route to the equation of state with simulation results is fairly good \cite{Miller2003} but can be improved by using the energy route to the equation of state \cite{Watts1971}. 

For our present purposes we note that the existence of an analytical solution for $c(r)$ allows us to obtain the FW line efficiently. The integrations in Eqs.~\eqref{Eq:DefPolesRealSpace_1} and \eqref{Eq:DefPolesRealSpace_2} can be performed analytically, leading to a system of three coupled algebraical equations to be solved for the decay length, the wavelength of the oscillations and the stickiness parameter $\tau$. The Widom line is also readily obtained from the equation for the correlation length $\xi = (\alpha_0^\text{mon})^{-1}$ of the monotonic solution.
\begin{figure}[t] 
	\centering
	\includegraphics[width = 8cm]{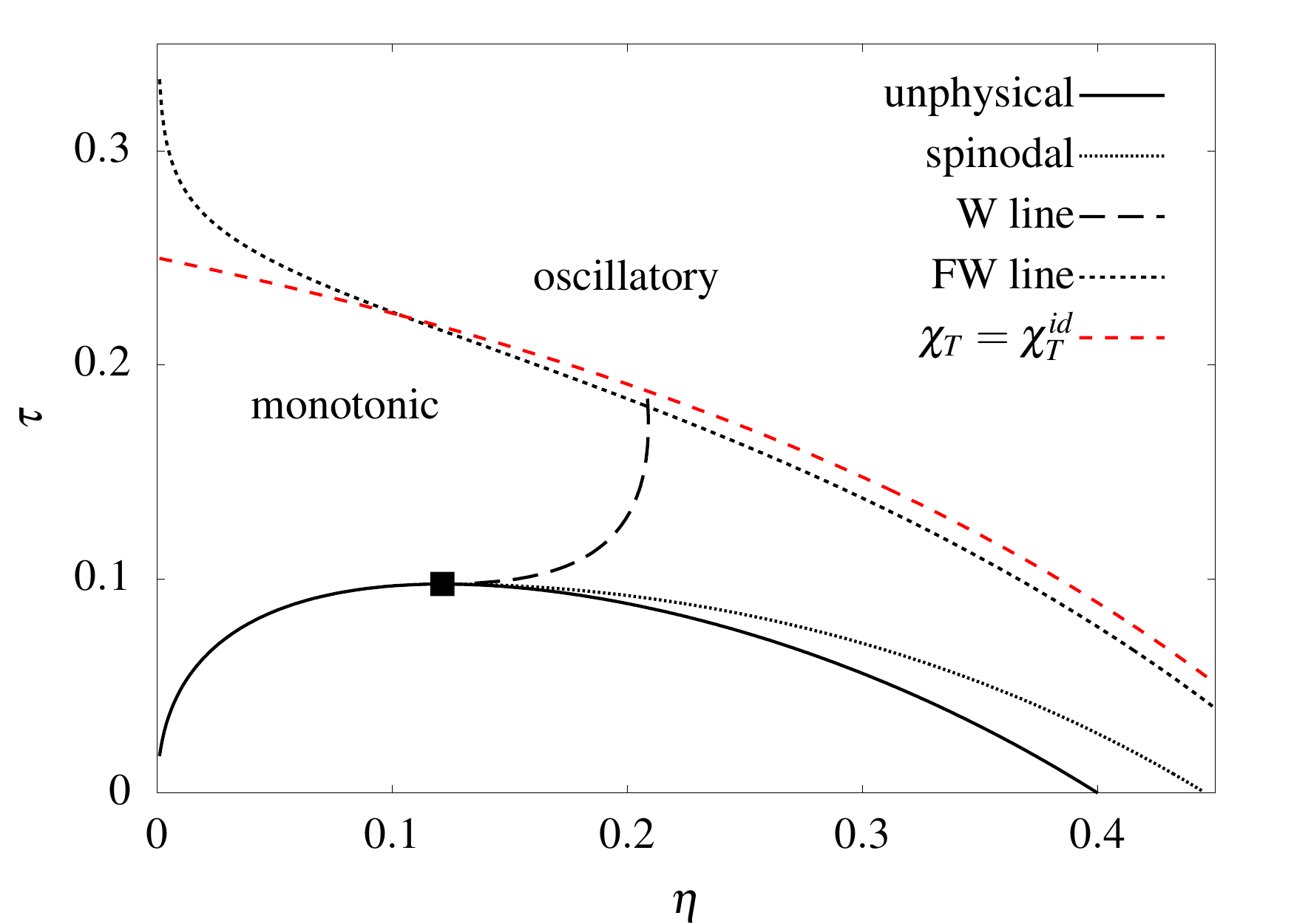}
	\caption{Phase diagram of the SHS fluid as obtained from Baxter's analytical solution \cite{Baxter1968} of the PY approximation plotted in the $\tau$-$\eta$ plane. The lines displayed correspond to those in Fig. \ref{Fig:phase_diag_SW} for SW, with the ordinate equivalent to temperature. The critical point (black square) is located at $(\eta_c, \tau_c) = (0.12, 0.09)$. The solid line denotes the region where the solution becomes unphysical. \red{Note that the W line terminates at the FW line as in Fig. \ref{Fig:phase_diag_AO}.}} \label{Fig:phase_diag_SHS}
\end{figure} 

We display the results in a $\tau$-$\eta$ phase diagram in Fig. \ref{Fig:phase_diag_SHS}. The FW line lies close to the  $\chi_T=\chi_T^{\text{id}}$ line throughout the region of physical interest. As $\eta\rightarrow 0$ the FW line increases rapidly, consistent with the asymptotic behavior discussed for the AO model in Sec. \ref{SubSec:AO}. As in the case of the SW fluid, there is an intersection between the two lines but now this occurs for a density smaller than the critical value.

The Widom line (long-dashed) is curious. It appears to approach the critical point horizontally and unlike in the SW and AO models, the W line shifts to higher \lq temperature\rq\, $\tau$ as $\eta$ is increased. However, these features might be an artifact associated by the limitations of the PY solutions for the SHS model. We note that for state points \red{inside the region bounded by the solid line} there are no-solution regions so that the complete spinodal cannot be determined.

 \subsection{The hard-core Yukawa fluid} \label{SubSec:MSA}
 
 \begin{figure}[t] 
 	\centering
 	\includegraphics[width = 8cm]{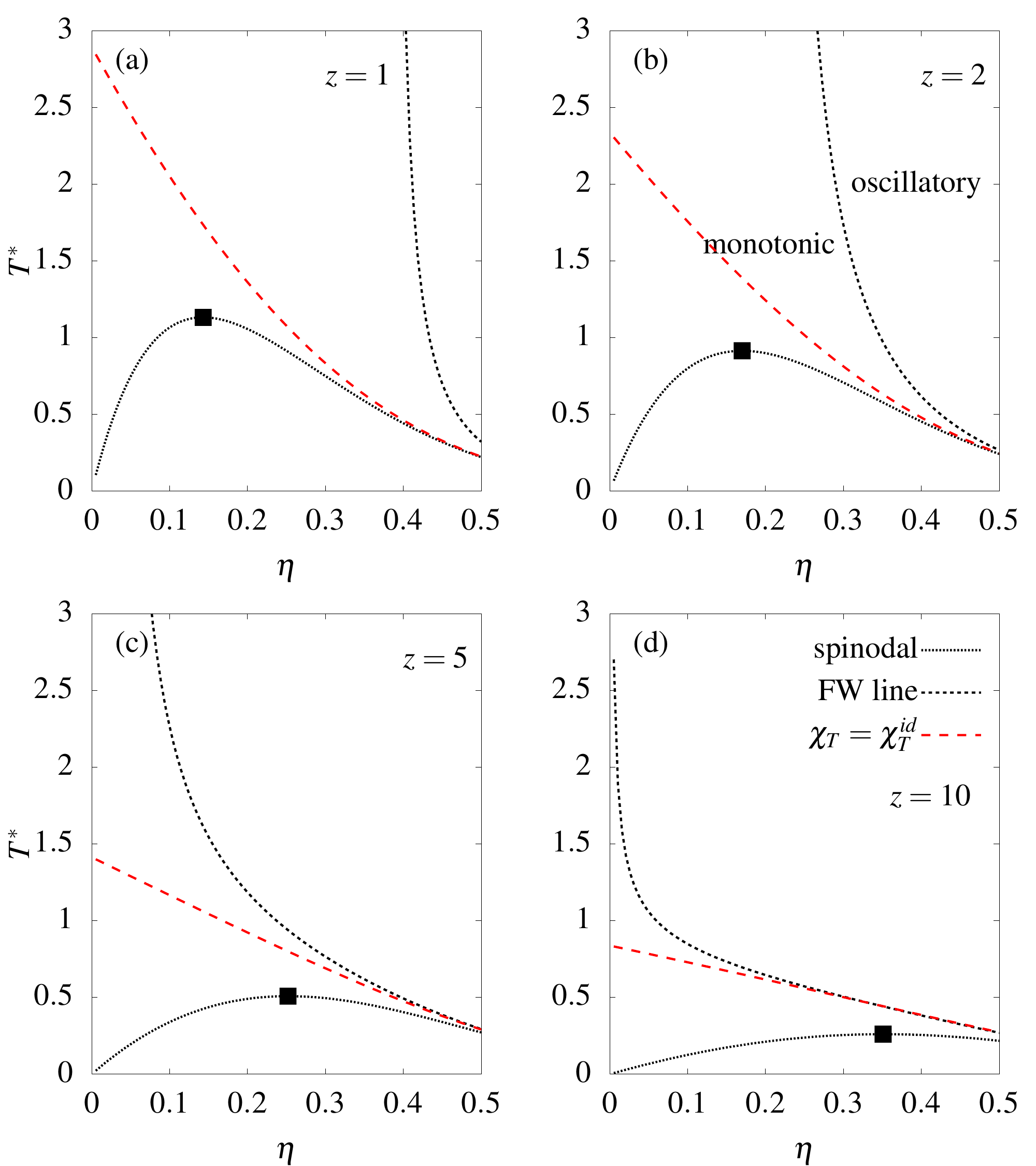}
 	\caption{Phase diagram of the Yukawa fluid treated within mean-spherical approximation \cite{Waisman1973} for  $z = 1$ (a), $2$ (b), $5$ (c), and 10 (d). The lines correspond to those in Fig. \ref{Fig:phase_diag_SW}. For clarity we do not display the binodal and W lines. The black square denotes the critical point. } \label{Fig:phase_diag_MSA}
 \end{figure} 

  The attractive hard-core Yukawa pair potential is given by
  \begin{equation}
   \phi_{\text{\tiny YUK}}(r) = 
  \begin{cases}
  \infty & r<\sigma \\
  -\frac{K e^{-z (r/\sigma-1)}}{r/\sigma} & r>\sigma\,,
  \end{cases}
  \end{equation}
  where the dimensionless parameter $z$ controls the range and $K>0$ measures the strength of the attraction. There is an analytical expression for $c(r)$ within the mean spherical approximation (MSA) due to Waisman \cite{Waisman1973}. From the MSA solution we calculate the FW and the $\chi_T^{\text{ex}} = 0$ lines as well as the spinodals for different values of $z$. We use the reduced temperature $T^*$ as introduced in Ref.~\onlinecite{Brown1996}. This sets
  \begin{equation}
  T^* =  k_B T \left[\frac{8}{9} \left(\frac{1}{3}+\frac{1}{z}+\frac{1}{z^2} \right) K\right]^{-1} \,,
  \end{equation}
  a choice that follows from mapping to the SW model within the framework of the MF approach of Sec. \ref{SubSec:SW}; see Ref. \onlinecite{Brown1996} for details.
  
  We display results for $z = 1$, 2, 5, and 10 in Figs. \ref{Fig:phase_diag_MSA} (a) -- (d), respectively. The key is the same \textcolor{black}{as} in previous figures. However, for the sake of clarity we do not show the binodals and the W lines. The corresponding critical points (black squares) are located at $(\eta_c, T_c^*) = (0.143, 1.13)$, (0.170, 0.914), (0.251, 0.508), and (0.350, 0.260).  The FW line for the case $z=1$ was calculated in Ref. \onlinecite{Brown1996} using the MSA and RPA; for this choice FW crossover occurs at very high values of $\eta$, which probably lie close to or beyond the freezing transition. With increasing $z$, i.e. decreasing range of the attraction, the critical density shifts to higher $\eta$ and the FW line to smaller $\eta$. For short-ranged attraction, i.e. large $z$, the FW and the $\chi_T^{\text{ex}} = 0$ lines lie very close together with deviations occurring only in the low-density regime. This is similar to the results for the other model fluids. Note that for $z\gtrsim 6$ the gas-liquid phase transitions becomes metastable with respect to the fluid-solid transition \cite{Hagen1994,Tuinier2006}. When the attraction becomes longer-ranged, i.e. $z \lesssim 5$, the FW and $\chi_T^{\text{ex}} = 0$ lines start to deviate significantly for all but the largest packing fractions. We speculate on the origin of this behavior in Sec. \ref{Sec:Conclusion}. In Fig. \ref{Fig:phase_diag_MSA} (d) we observe the FW line approaching asymptotically the $T^*$ axis as $\eta\rightarrow 0$. Once again this is consistent with general considerations \cite{Dijkstra2000, Roth2018}.
 
 \section{Discussion and Conclusion} \label{Sec:Conclusion}
 
 We began our paper by recalling the importance of the second virial coefficient $B_2(T)$ for quantifying the balance between net repulsion and attraction that determines certain thermodynamic properties of fluids. The Boyle temperature $T_B$ represents the simplest example of a crossover temperature identified by the vanishing of the excess pressure, $p-p^\text{id}$, in a dilute gas. Here we have identified a more sophisticated criterion, pertinent to a dense fluid, that also reflects the competition between repulsive and attractive interactions. The new criterion also involves a thermodynamic excess function: the line in the phase diagram where the excess isothermal compressibility $\chi_T^\text{ex}$ vanishes describes crossover of the total pair correlation function $rh(r)$ from asymptotic damped oscillatory decay, dominated by repulsive forces and characteristic of HS systems, to monotonic decay, dominated by attractive forces and characteristic of dilute gaseous and near-critical fluid states. Our criterion corresponds to the condition $(\partial \beta p / \partial \rho_b)_T = 1$. In the limit $\rho_b\rightarrow 0$ this condition reduces to $B_2(T_B)=0$, i.e. the definition of the Boyle temperature; \textcolor{black}{however note that approximate theories do not necessarily yield the exact Boyle temperature}. Generally, the condition corresponds to the static structure factor $S(k)$, at wave number $k=0$,  taking its ideal gas value: $S(0) =1$. For four distinct types of model fluid this new thermodynamic criterion yields a line in the $T-\rho$ phase diagram that lies close to the actual FW crossover line, as determined from a formal pole analysis of the asymptotic decay \cite{FisherWidom1968, HansenMcDonald2013, Evans1993, Evans1994}. 
 
 The new criterion is based on considering the attractive and repulsive contributions to the pair direct correlation function $c(r)$, see Sec.~\ref{Sec:ParticleForcesAndCompr}, for a range of thermodynamic states. It corresponds to the condition $C=0$, i.e. the three dimensional integral of $c(r)$ in Eq.~\eqref{Eq:Def_C}, vanishes. For the simple square well (SW) model the exact asymptotic result $c(r) = -\beta\phi(r)$ describes well the behavior of $c(r)$ outside the hard-core (see Fig.~\ref{Fig:cr_MC}). It is then clear how the condition $C=0$ arises and the origin of the crossover from monotonic to damped oscillatory as the density increases is straightforward to grasp. Unfortunately, there are few simulation results in the literature, for models exhibiting both repulsive and attractive interactions, that display $c(r)$ at different state points. Ref.~\onlinecite{Dijkstra2000} for the truncated and shifted Lennard-Jones model is an exception and it is pleasing that the MC results for a fixed supercritical temperature also show the asymptotic formula for $c(r)$ holding outside the core region and the core contribution becoming more negative as the density increases. Without knowledge of $c(r)$ one cannot perform a careful pole analysis of the asymptotic decay required to determine the true correlation length and therefore the FW and W lines. One must resort to fitting simulation data for the decay of $\ln [r h(r)]$ to obtain decay lengths.
 
 From our present results we cannot say how reliable the new criterion $\chi_T=\chi_T^\text{id}$ is for a general pair potential exhibiting repulsion and attraction. Our results point to its reliability for short-ranged potentials. For the AO model considered in Sec.~\ref{SubSec:AO} the size-ratio $q=0.6$ corresponds to intermediate ranged effective colloid-colloid interactions. For the shorter ranged case, $q= 0.3$, the agreement between the two relevant lines is equally good. Of course, the sticky HS fluid in Sec.~\ref{SubSec:SHS} is certainly short-ranged. More complex is the hard-core Yukawa model treated in Sec.~\ref{SubSec:MSA}. The new criterion works well for large $z$, a rapidly decaying Yukawa. However, for smaller $z$ we observe considerable deviations between the FW and the $\chi_T=\chi_T^\text{id}$ lines. Within the MSA the FW line is pushed further towards higher densities and the onset of the crystal as $z$ decreases. This leads us to wonder how reliable the MSA prediction is for longer-ranged potentials. It might well be that the criterion $C=0$ is no longer an accurate approximation for the FW line. Once again, the form of $c(r)$ is important and one needs to have a reliable theory for this as the range of the potential increases.
 
 The law of corresponding states presented in Appendix \ref{Sec:Appendix_A} describes the line $\chi_T=\chi_T^\text{id}$ in terms of temperature and density reduced w.r.t. their respective critical parameters. The result \eqref{Eq:CorrespondingStates} is derived for the mean-field treatment of attractive interactions, i.e. for fluids described by a generalized van der Waals equation of state \eqref{Eq:EquationOfState_SW}. Nevertheless, we expect the same form to remain reasonably accurate for other models and equations of states.
 
 We calculated the Widom W line for each model fluid and in Figs.~\ref{Fig:phase_diag_AO} and \ref{Fig:phase_diag_SHS} we show this line terminating at the FW line. As explained in Sec.~\ref{Sec:Results}, the oscillatory pole has no minimum as a function of density or packing fraction of the colloids. Therefore, the W line, which we define as the line of a local maximum of the \textit{true} correlation length $\xi$, is restricted to the region of the phase diagram where monotonic decay pertains. Of course, crossing the FW line does not signal any singular behavior of thermodynamic response functions since the latter always involve integrals over $h(r)$ which include contributions from both the leading and next to leading (plus higher-order) poles. Note that if we were to define the W line as the locus of local maxima of the OZ correlation length $\xi_{OZ}$, as in Ref.~\onlinecite{Brazhkin2014}, this line would not reflect the pole structure, i.e. the true asymptotic decay of the pair correlation function \cite{Footenote}. Rather it would reflect primarily the extrema in $S (0)$ -- see Eq.~\eqref{Def:xi_OZ}.
 
 The shape of our W lines in the $T - \rho$ plane is interesting. The sticky-sphere result in Fig.~\ref{Fig:phase_diag_SHS} has a very different form from the SW and AO results. Interestingly, the W line determined for sticky-spheres is similar to that calculated in a recent study of the Jagla model \cite{Lopez2018} -- see their Fig.~8: the W line emanates from the critical point and shifts to higher $\rho$ as $T$ increases, then rises near vertically to the FW line. The authors show the W line extending into the oscillatory decay region as $T$ increases but are careful to point out that there is no maximum of the true correlation length once the FW line is crossed. Their extension corresponds to the locus of minima of $\alpha_0^\text{mon}$. 
 
 We should recall that FW lines also occur in dimensions $d < 3$. Indeed, the original FW paper \cite{FisherWidom1968} was based on exact results for one-dimensional $(d=1)$  models. Recent studies for SW \cite{Fantoni2017} and for triangle-well potentials \cite{Montero2019} in $d=1$ provide exact results for the FW line. It would be instructive to compare these with those based on our new criterion $\chi_T=\chi_T^\text{id}$. We shall return to this in a future paper. \textcolor{black}{It is also interesting to note that the W lines shown in Ref.~\onlinecite{Montero2019} have a similar shape as those in the three-dimensional SW and AO models, though these extend down to zero temperature as there is no critical point in $d = 1$.}
 
 As a final remark we note that it is difficult to determine the FW and the W line in simulations because the true correlation length $\xi$ is not easily computed. It requires knowledge of $h(r)$ with high precision out to large separations $r$, i.e. to ten particle diameters or more, to be able to make reliable fits to the decay lengths or to perform an accurate determination of $c(r)$ in order to determine poles. The $\chi_T =\chi_T^\text{id}$ criterion is much easier to examine in simulations. As such this is a good starting point for investigating where in the phase diagram we expect the competition between repulsive and attractive interactions to lead to crossover in the asymptotic decay of correlations, i.e. to point to the location of a FW line. Experimentally $\chi_T$ is easily obtained from equation of state determinations or from extrapolation of small angle scattering data to zero wavenumber \cite{HansenMcDonald2013}.
 
\section*{Acknowledgements}

D. S. acknowledges financial support from the Carl-Zeiss-Stiftung and the hospitality of the Theoretical Physics Group at the University of Bristol during his visit.

 \appendix
 
 \section{The line $\chi_T = \chi_T^\text{id} $ from the mean-field treatment of attraction} \label{Sec:Appendix_A}
 
 For a general attractive potential the pressure resulting from the mean-field approach is given by
 \begin{equation} \label{Eq:EoSGeneral}
 	p = p_\text{\tiny CS} + \frac{\rho_b^2}{2} \,\widehat{\phi}_\text{att}(0)\,,
 \end{equation}
 where $p_\text{\tiny CS}$ is the Carnahan-Starling (CS) result for hard-spheres, given by the first term in Eq. \eqref{Eq:EquationOfState_SW}. At the critical point $(\partial p/\partial \rho_b)_{\rho_c, T_c} = 0$ so that
 \begin{equation} \label{Eq:PressureCritPoint}
 	\left(\frac{\partial p_\text{\tiny CS}}{\partial \rho_b}\right)_{\rho_c, T_c} = -\rho_c\,\widehat{\phi}_\text{att}(0)\,.
 \end{equation}
 But the condition $\chi_T = \chi_T^\text{id}$ yields the line $\beta (\partial p/\partial \rho_b)_T = 1$, or
 \begin{equation} \label{Eq:PressureStopper}
 	\left(\frac{\partial p_\text{\tiny CS}}{\partial \rho_b}\right)_{\rho_b, T} + \rho_b\, \widehat{\phi}_\text{att}(0) = k_B T\,.
 \end{equation}
Eliminating $\widehat{\phi}_\text{att}(0)$ between Eqs. \eqref{Eq:PressureCritPoint} and \eqref{Eq:PressureStopper} we find
\begin{equation} \label{Eq:PressureNoPotential}
	\rho_b\,\left(\frac{\partial p_\text{\tiny CS}}{\partial \rho_b}\right)_{\rho_c, T_c} - \rho_c\left(\frac{\partial p_\text{\tiny CS}}{\partial \rho_b}\right)_{\rho_b, T} = - k_B T \rho_c\,.
\end{equation}
Noting that the CS pressure can be written as $p_\text{\tiny CS} \equiv k_B T \rho_b\, Z_\text{\tiny CS}(\eta)$, and that other hard-sphere equations of state take the same form, we can re-write Eq. \eqref{Eq:PressureNoPotential} as
\begin{equation} \label{Eq:CorrespondingStates}
	\frac{\eta}{\eta_c} \frac{T_c}{T} I(\eta_c) = I(\eta)-1\,,
\end{equation}
where 
\begin{equation} \label{Eq:CorrespondingStates_2}
	I(\eta)\equiv Z_\text{\tiny CS}(\eta)+\eta (\partial Z_\text{\tiny CS}(\eta)/\partial \eta)
\end{equation}
 is easily determined. It is well-known that the critical value of the packing fraction $\eta_c$, obtained from the equation of state \eqref{Eq:EoSGeneral} is independent of $\widehat{\phi}_\text{att}(0)$. Within CS one finds $\eta_c = 0.1304$. Clearly, Eq. \eqref{Eq:CorrespondingStates} takes the form of a law of corresponding states for the $T/T_c$ vs. $\eta/\eta_c$ line defined by $\chi_T = \chi_T^\text{id}$. Using a hard-sphere equation of state different from CS will alter $Z(\eta)$ but the form of \eqref{Eq:CorrespondingStates} and \eqref{Eq:CorrespondingStates_2} is retained.

\bibliographystyle{apsrev4-1}
\bibliography{references}

\end{document}